\documentclass[preprint]{elsarticle}  % for review and submission

\usepackage{graphicx}  % needed for figures
\usepackage{graphics}
\usepackage{dcolumn}   % needed for some tables
\usepackage{bm}        % for math
\usepackage{amssymb}   % for math
\usepackage{amsmath}   % for math
\usepackage{caption}
\usepackage{epsfig}

\begin{document}
\begin{frontmatter}

\title{Electroexcitation of the $\Delta^{+}(1232)$ at low momentum transfer}

\author[temple]{A.~Blomberg}
\address[temple]{Temple University, Philadelphia, PA 19122, USA}

\author[dalhousie,saintmary]{D.~Anez}
\address[dalhousie]{Dalhousie University, Halifax, Nova Scotia, Canada}
\address[saintmary]{Saint Mary's University, Halifax, Nova Scotia, Canada}

\author[temple]{N.~Sparveris\corref{cor1}}
\ead{sparveri@temple.edu}
\cortext[cor1]{Corresponding author.}

\author[saintmary]{A.~J.~Sarty}

\author[temple]{M.~Paolone}

\author[mit]{S.~Gilad}
\address[mit]{Massachusetts Institute of Technology, Cambridge, MA
02139, USA}

\author[jlab]{D.~Higinbotham}
\address[jlab]{Thomas Jefferson National Accelerator Facility, Newport
News, VA 23606, USA}

\author[syra]{Z.~Ahmed}
\address[syra]{Syracuse University, Syracuse, NY 13210, USA}

\author[odu]{H.~Albataineh}
\address[odu]{Old Dominion University, Norfolk, VA 23529, USA}

\author[jlab]{K.~Allada}

\author[kent]{B.~Anderson}
\address[kent]{Kent State University, Kent, OH 44242, USA}

\author[cal]{K.~Aniol}
\address[cal]{California State University, Los Angeles, Los Angeles, CA 90032, USA}

\author[glasgow]{J.~Annand}
\address[glasgow]{University of Glasgow, Glasgow G12 8QQ, Scotland, UK}

\author[argonne]{J.~Arrington}
\address[argonne]{Physics Division, Argonne National Laboratory, Argonne,
IL 60439, USA}

\author[wm]{T.~Averett}
\address[wm]{The College of William and Mary, Williamsburg, VA 23187, USA}

\author[uva]{H.~Baghdasaryan}
\address[uva]{University of Virginia, Charlottesville, VA 22904, USA}

\author[ciae]{X.~Bai}
\address[ciae]{China Institute of Atomic Energy, Beijing, China}

\author[negev]{A.~Beck}
\address[negev]{Nuclear Research Center-Negev, Beer-Sheva, Israel}

\author[negev]{S.~Beck}

\author[catania]{V.~Bellini}
\address[catania]{Universita di Catania , Catania, Italy}

\author[duquesne]{F.~Benmokhtar}
\address[duquesne]{Duquesne University, Pittsburgh, PA 15282, USA}

\author[fia]{W.~Boeglin}
\address[fia]{Florida International University, Miami, FL 33199, USA}

\author[ipn]{C.~M.~Camacho}
\address[ipn]{Institut de Physique Nucleaire, Orsay, France}

\author[jlab]{A.~Camsonne}

\author[hampton]{C.~Chen}
\address[hampton]{Hampton University , Hampton, VA 23668, USA}

\author[jlab]{J.~P.~Chen}

\author[uva,chiang]{K.~Chirapatpimol}
\address[chiang]{Chiang Mai University, Chiang Mai, Thailand}

\author[infn]{E.~Cisbani}
\address[infn]{INFN, Sezione di Roma, I-00161 Rome, Italy}

\author[uva]{M.~Dalton}

\author[wm]{W.~Deconinck}

\author[ecole]{M.~Defurne}
\address[ecole]{\'Ecole Centrale Paris, Ch\^atenay-Malabry, France}

\author[bari]{R.~De Leo}
\address[bari]{Universite di Bari, Bari, Italy}

\author[temple]{D.~Flay}

\author[utenn]{N.~Fomin}
\address[utenn]{University of Tennessee, Knoxville, TN 37996, USA}

\author[cmu]{M.~Friend}
\address[cmu]{Carnegie Mellon University, Pittsburgh, PA 15213, USA}

\author[infn,iss]{S.~Frullani}
\address[iss]{Istituto Superiore di Sanit\`a, I-00161 Rome, Italy}

\author[temple]{E.~Fuchey}

\author[infn]{F.~Garibaldi}

\author[rutgers]{R.~Gilman}
\address[rutgers]{Rutgers University, New Brunswick, NJ 08855, USA}

\author[uva]{C.~Gu}

\author[glasgow]{D.~Hamilton}

\author[uva,fsu]{C.~Hanretty}
\address[fsu]{Florida State University, Tallahassee, FL 32306, USA}

\author[jlab]{O.~Hansen}

\author[uva]{M.~Hashemi Shabestari}

\author[telaviv]{O.~Hen}
\address[telaviv]{Tel Aviv University, Tel Aviv 69978, Israel}

\author[longwood]{T.~Holmstrom}
\address[longwood]{Longwood University, Farmville, VA 23909, USA}

\author[duke]{M.~Huang}
\address[duke]{Duke University, Durham, NC 27708, USA}

\author[cal]{S.~Iqbal}

\author[houston]{N.~Kalantarians}
\address[houston]{Unversity of Houston, Houston, TX 77030, USA}

\author[seoul]{H.~Kang}
\address[seoul]{Seoul National University, Seoul, Korea}

\author[bates]{A.~Kelleher}
\address[bates]{MIT Bates Linear Accelerator, Middleton, MA 01949, USA}

\author[indiana]{M.~Khandaker}
\address[indiana]{Indiana University , Bloomington, IN 47405, USA}

\author[telaviv]{I.~Korover}

\author[indiana]{J.~Leckey}

\author[jlab]{J.~LeRose}

\author[uva]{R.~Lindgren}

\author[kent]{E.~Long}

\author[vpi]{J.~Mammei}
\address[vpi]{Virginia Polytechnic Institute and State University, Blacksburg, VA 24061, USA}

\author[cal]{D.J.~Margaziotis}

\author[corp]{A.~Mart\'i Jimenez-Arguello}
\address[corp]{Laboratoire de Physique Corpusculaire de Clermont-Ferrand, Aubière Cedex, France}

\author[jlab]{D.~Meekins}

\author[temple]{Z.~E.~Meziani}

\author[jsi]{M.~Mihovilovic}
\address[jsi]{Jo\v{z}ef Stefan Institute, Ljubljana, Slovenia}

\author[bates]{N.~Muangma}

\author[uva]{B.~Norum}

\author[miss]{Nuruzzaman}
\address[miss]{Mississippi State University, Mississippi State, MS
39762, USA}

\author[mit]{K.~Pan}

\author[unh]{S.~Phillips}
\address[unh]{University of New Hampshire, Durham, NH 03824, USA}

\author[telaviv]{E.~Piasetzky}

\author[temple]{A.~Polychronopoulou}

\author[telaviv]{I.~Pomerantz}

\author[temple]{M.~Posik}

\author[norfolk]{V.~Punjabi}
\address[norfolk]{Norfolk State University, Norfolk, VA 23504, USA}

\author[duke]{X.~Qian}

\author[sns]{A.~Rakhman}
\address[sns]{Spallation Neutron Source, Oak Ridge National Laboratory,
Oak Ridge, TN 37831, USA}

\author[argonne]{P.~E.~Reimer}

\author[umass]{S.~Riordan}
\address[umass]{University of Massachusetts, Amherst, MA 01003, USA}
\address[stony]{Stony Brook University, Stony Brook, NY 11794, USA}

\author[racah]{G.~Ron}
\address[racah]{Racah Institute of Physics, Hebrew University of
Jerusalem, Jerusalem, Israel 91904}

\author[jlab]{A.~Saha}

\author[temple]{E.~Schulte}

\author[kent]{L.~Selvy}

\author[telaviv]{R.~Shneor}

\author[ul]{S.~Sirca}
\address[ul]{University of Ljubljana, Ljubljana, Slovenia}

\author[glasgow]{J.~Sjoegren}

\author[gwu]{R.~Subedi}
\address[gwu]{George Washington University, Washington, DC 20052,
USA}

\author[jlab]{V.~Sulkosky}

\author[mich]{W.~Tireman}
\address[mich]{Northern Michigan University, Marquette, MI 49855, USA}

\author[uva]{D.~Wang}

\author[kent]{J.~Watson}

\author[jlab]{B.~Wojtsekhowski}

\author[hefei]{W.~Yan}
\address[hefei]{University of Science and Technology of China, Hefei
230026, People's Republic of China}

\author[telaviv]{I.~Yaron}

\author[uva]{Z.~Ye}

\author[argonne]{X.~Zhan}

\author[jlab]{J.~Zhang}

\author[rutgers]{Y.~Zhang}

\author[ct]{B.~Zhao}
\address[ct]{University of Connecticut, Storrs, CT 06269, USA}

\author[uva]{Z.~Zhao}

\author[uva]{X.~Zheng}

\author[hefei]{P.~Zhu}

%\date{\today}
\begin{abstract}
We report on new p$(e,e^\prime p)\pi^\circ$ measurements at the
$\Delta^{+}(1232)$ resonance at the low momentum transfer region,
where the mesonic cloud dynamics is predicted to be dominant and
rapidly changing, offering a test bed for chiral effective field
theory calculations. The new data explore the $Q^2$ dependence of
the resonant quadrupole amplitudes and for the first time indicate
that the Electric and the Coulomb quadrupole amplitudes converge as
$Q^2\rightarrow0$. The measurements of the Coulomb quadrupole
amplitude have been extended to the lowest momentum transfer ever
reached, and suggest that more than half of its magnitude is
attributed to the mesonic cloud in this region. The new data
disagree with predictions of constituent quark models and are in
reasonable agreement with dynamical calculations that include pion
cloud effects, chiral effective field theory and lattice
calculations. The measurements indicate that improvement is required
to the theoretical calculations and provide valuable input that will
allow their refinements.
\end{abstract}

\begin{keyword}
CMR \sep pion electroproduction \sep nucleon deformation

\PACS 13.60.Le \sep 13.40.Gp \sep 14.20.Gk
\end{keyword}

\end{frontmatter}

%=====================================
% Introduction
%=====================================

%\section{Introduction}

The $\Delta$(1232) resonance - the first excited state of the
nucleon - dominates many nuclear phenomena at energies above the
pion-production threshold and plays a prominent role in the physics
of the strong interaction. The study of the $\Delta$ has allowed to
explore various aspects of the nucleonic structure, such as the
study of d-wave components that could quantify to what extent the
nucleon or the $\Delta$ wave function deviates from the spherical
shape~\cite{amb}, or more recently the exploration of the
Generalized Polarizabilities (GPs) of the nucleon which, contrary to
the elastic form factors, are sensitive to all the excited spectrum
of the nucleon~\cite{jlabgp,gpprop,mami2}.

Hadrons are composite systems with complex quark-gluon and meson
cloud dynamics that give rise to non-spherical components in their
wavefunction, which in a classical limit and at large wavelengths
will correspond to a ``deformation"~\cite{Ru75,glashow,soh}. The
determination and subsequent understanding of the shapes of the
fundamental building blocks in nature is a particularly fertile line
of investigation for the understanding of the interactions of their
constituents amongst themselves and the surrounding medium. For
hadrons this means the interquark interaction and the quark-gluon
dynamics. For the proton, the only stable hadron, the vanishing of
the spectroscopic quadrupole moment, due to its spin 1/2 nature,
precludes access to the most direct observable of deformation. As a
result, the presence of the resonant quadrupole amplitudes
$E^{3/2}_{1+}$ and $S^{3/2}_{1+}$ (or E2 and C2 photon absorption
multipoles respectively) in the predominantly magnetic dipole
$M^{3/2}_{1+}$ (or M1) $\gamma^* N\rightarrow \Delta$ transition has
emerged as the experimental signature for such an effect
\cite{amb,soh,Ru75,glashow,glas2,capstick,pho2,pho1,pho1b,frol,pos01,merve,bart,Buuren,joo,spaprc,kun00,joo1,joo2,spaprl,kelly,stave,ungaro,oops,dina,dina2,dina3,sato_lee,dmt,kama,mai00,multi,said,elsner,spaplb,longpaper,aznauryan,villano,kirkpatrick,sparepja,vcspaper,quarkpion1,quarkpion2,quarkpion3,pasc,hemmert,hqm,mande}.
%%The spectroscopic quadrupole moment provides the most reliable and
%%interpretable measurement of such amplitudes. For the proton it
%%vanishes identically because of its spin 1/2 nature. Instead, the
%%signature of the non-spherical components of the proton is sought in
%%the presence of resonant quadrupole amplitudes
%%$(E^{3/2}_{1+}, S^{3/2}_{1+})$ in the predominantly magnetic dipole
%%($M^{3/2}_{1+}$) $\gamma^* N\rightarrow \Delta$ transition.
Nonvanishing quadrupole amplitudes will signify that either the
proton or the $\Delta^{+}(1232)$ or more likely both are
characterized by non-spherical components in their wavefunctions.
These amplitudes have been explored up to four momentum transfer
squared $Q^2=7~(GeV/c)^2$
\cite{pho2,pho1,pho1b,frol,pos01,merve,bart,Buuren,joo,spaprc,kun00,joo1,joo2,spaprl,kelly,stave,ungaro,elsner,spaplb,longpaper,aznauryan,villano,kirkpatrick,sparepja}.
Their relative strength is normally quoted in terms of the ratios
EMR $= Re(E^{3/2}_{1+}/M^{3/2}_{1+})$ and CMR $=
Re(S^{3/2}_{1+}/M^{3/2}_{1+})$. The experimental results are in
reasonable agreement with models invoking the presence of
non-spherical components in the nucleon wavefunction.

%-----------------------------------------------------------------------------
\begin{figure}[!]
\includegraphics[width = \columnwidth]{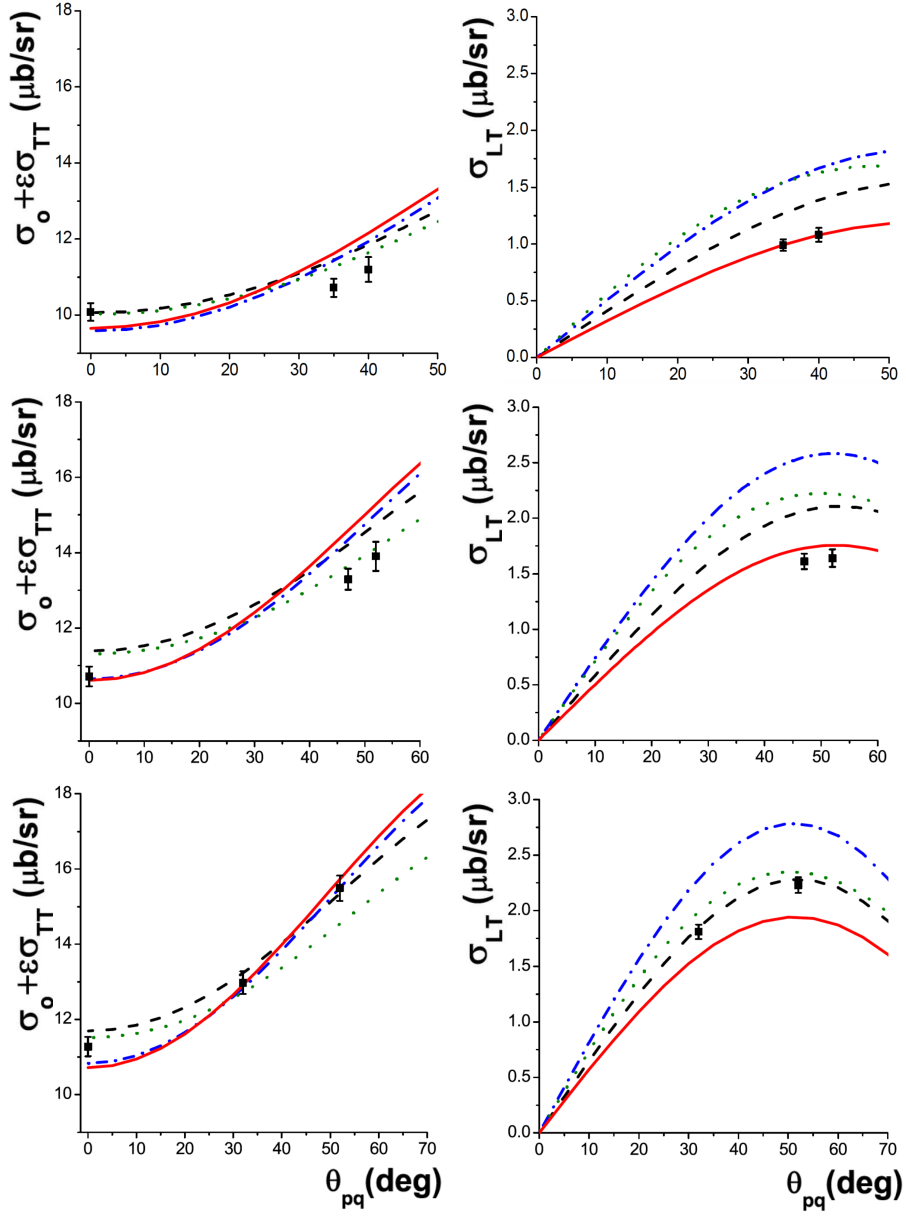}
%%\begin{figure*}
%%\centerline{\psfig{figure=resp.eps,width=12.0cm,angle=0.0}}
\smallskip
\caption{Measurements of $\sigma_{0}+\epsilon \cdot \sigma_{TT}$ and
$\sigma_{LT}$ at $Q^2 = 0.04~(GeV/c)^2$ (top panels), $Q^2 =
0.09~(GeV/c)^2$ (center), and $Q^2 = 0.13~(GeV/c)^2$ (bottom). The
theoretical predictions of DMT \cite{dmt} (dash-dot), SAID
\cite{said} (dot), MAID \cite{kama,mai00} (dash), and Sato Lee
\cite{sato_lee} (solid) are also presented.} \label{fig:resp}
%%\end{figure*}
\end{figure}
%-----------------------------------------------------------------------------

In the constituent-quark picture of hadrons, the non-spherical
amplitudes are a consequence of  the non-central, color-hyperfine
interaction among quarks \cite{glashow,glas2}. However, it has been
shown that this mechanism only provides a small fraction of the
observed quadrupole signal at low momentum transfers, with the
magnitudes of this effect for the predicted E2 and C2 amplitudes
\cite{capstick} being at least an  order of magnitude too small to
explain the experimental results and with the dominant M1 matrix
element being $\approx$~30\% low. A likely cause of these dynamical
shortcomings is that such quark models do not respect chiral
symmetry, whose spontaneous breaking leads to strong emission of
virtual pions (Nambu-Goldstone Bosons) \cite{amb}. These couple to
nucleons as $\vec{\sigma}\cdot \vec{p}$ where $\vec{\sigma}$ is the
nucleon spin, and $\vec{p}$ is the pion momentum. The coupling is
strong in the p wave and mixes in non-zero angular momentum
components. Based on this, it is physically reasonable to expect
that the pionic contributions increase the M1 and dominate the E2
and C2 transition matrix elements in the low $Q^2$ (large distance)
domain. This was first indicated by adding pionic effects to quark
models \cite{quarkpion1,quarkpion2,quarkpion3}, subsequently in pion
cloud model calculations \cite{sato_lee,dmt}, and recently
demonstrated in Chiral Effective Field Theory calculations
\cite{pasc}. With the existence of these non-spherical amplitudes
well established recent high precision experiments and theoretical
efforts have focused on testing in depth the reaction calculations
and decoding the underlying nucleon dynamics.

More recently the study of the $N\rightarrow \Delta$ transition has
emerged as an excellent testing ground to study the Generalized
Polarizabilities of the nucleon \cite{jlabgp,gpprop,mami2}. The GPs
are fundamental quantities of the nucleon. They can be seen as
Fourier transforms of local polarization densities (electric,
magnetic, and spin) allowing us to study the role of the pion cloud
and quark core contributions at various length scales. The
sensitivity to the GPs grows significantly in the resonance region,
and the precise knowledge of the $N\rightarrow \Delta$ transition
form factors is required as an input to Dispersions Relations
calculations \cite{gp12,gp13} in order to extract the GPs from
Virtual Compton Scattering measurements at the resonance region.

%\section{Results}

In this Letter we report on $\pi^\circ$~reaction channel
measurements at the low momentum transfer region. The new data
explore the $Q^2$ dependence of the quadrupole amplitudes with high
precision, and extend the measurements of the Coulomb quadrupole
amplitude to a new lowest momentum transfer. The cross section of
the $p(\vec{e},e^\prime p)\pi^\circ$ reaction is sensitive to five
independent partial responses ($\sigma_{T},\sigma_{L},\sigma_{LT},
\sigma_{TT}$ and $\sigma_{LT'}$) \cite{multi} :
\begin{eqnarray}
 \frac{d^5\sigma}{d\omega d\Omega_e d\Omega^{cm}_{pq}} & = & \Gamma (\sigma_{T} + \epsilon{\cdot}\sigma_L
  - v_{LT}{\cdot}\sigma_{LT}{\cdot}\cos{\phi_{pq}^{*}} \nonumber \\
 & &   +\epsilon{\cdot}\sigma_{TT}{\cdot}\cos{2\phi_{pq}^{*}} \\
 & &   - h {\cdot} p_e {\cdot} v_{LT'}{\cdot}\sigma_{LT'}{\cdot}\sin{\phi_{pq}^{*}}) \nonumber
\label{equ:cros}
\end{eqnarray}
where  $v_{LT}=\sqrt{2\epsilon(1+\epsilon)}$ and
$v_{LT'}=\sqrt{2\epsilon(1-\epsilon)}$ are kinematic factors,
$\epsilon$ is the transverse polarization of the virtual photon,
$\Gamma$ is the virtual photon flux, $h= \pm 1$ is the electron
helicity, $p_e$ is the magnitude of the electron longitudinal
polarization, and $\phi_{pq}^{*}$ is the proton azimuthal angle with
respect to the electron scattering plane. The  differential cross
sections ($\sigma_{T},\sigma_{L},\sigma_{LT}, \sigma_{TT}$, and
$\sigma_{LT'}$) are all functions of the center-of-mass energy W,
the $Q^2$, and the proton center of mass polar angle
$\theta_{pq}^{*}$ (measured from the momentum transfer direction)
\cite{multi}. The $\sigma_{0}=\sigma_{T}$ +
$\epsilon\cdot\sigma_{L}$ response is dominated by the $M_{1+}$
resonant multipole while the interference of the $C2$ and $E2$
amplitudes with the $M1$ dominates the Longitudinal~-~Transverse and
Transverse~-~Transverse responses, respectively.

Measurements were made in Hall A at Jefferson Lab. A 15~$\mu A$ to
80~$\mu A$, 1160 MeV electron beam impinged on a 4~cm
liquid-hydrogen target. Electrons and protons were detected in
coincidence with the two High Resolution Spectrometers (HRS)
\cite{hrs}. Both spectrometers employ a pair of vertical drift
chambers for track reconstruction, three scintillator panels for
trigger, timing, and detector efficiencies, as well as two layers of
lead glass calorimeters. The electron spectrometer utilized a gas
Cherenkov detector. Both spectrometers are characterized by a
momentum resolution of $10^{-4}$ and a spectrometer angle
determination accuracy of 0.1~mr.

\begin{figure}[!]
\center
\includegraphics[width = \columnwidth]{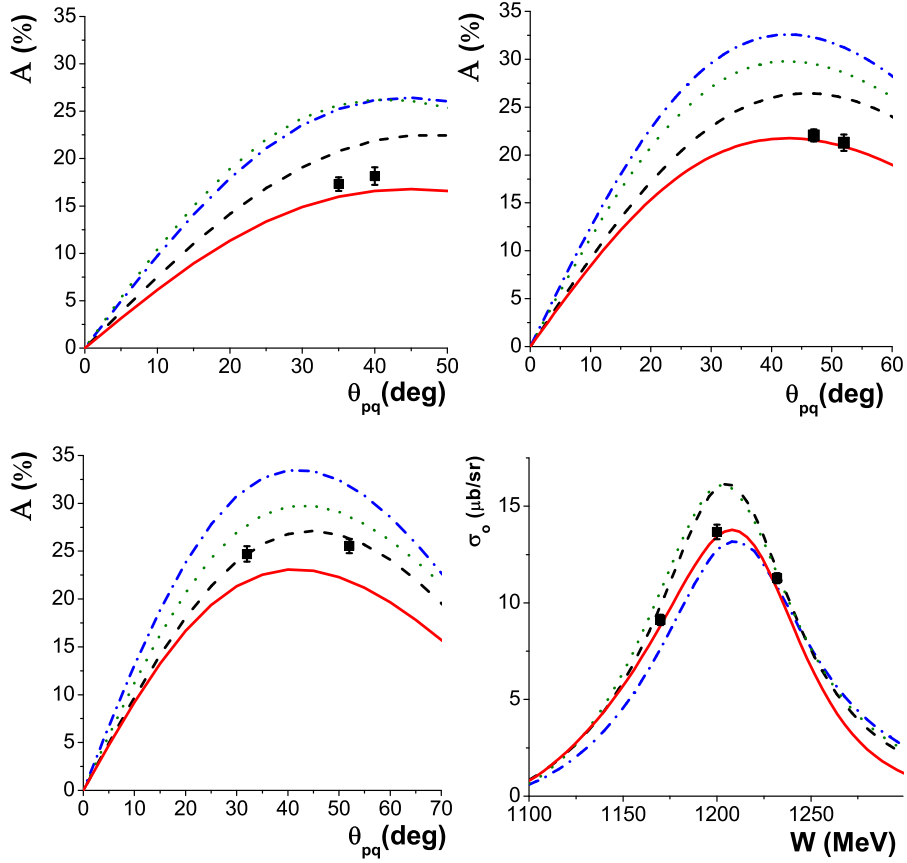}
\caption{Top panels: asymmetries at $Q^2 = 0.04~(GeV/c)^2$ (left)
and $Q^2 = 0.09~(GeV/c)^2$ (right). Bottom panels: asymmetries
(left) and $\sigma_{0}$ (right) at $Q^2 = 0.13~(GeV/c)^2$. The
definition of the theoretical curves is given at the caption of
Fig.~\ref{fig:resp}.} \label{fig:asym}
\end{figure}

Measurements were performed from $Q^2 = 0.04$ to $Q^2 =
0.13~(GeV/c)^2$. For each $\theta_{pq}^{*}$ setting the proton
spectrometer was sequentially placed at $\phi_{pq}^{*}= 0^\circ$ and
$180^\circ$, thus allowing to extract the $\sigma_{LT}$ and the
$\sigma_{0}+\epsilon \cdot \sigma_{TT}$ responses. The in-plane
azimuthal asymmetry of the cross section with respect to the
momentum transfer direction, $A_{(\phi_{pq}=0,\pi)} =
[\sigma_{\phi_{pq}=0} - \sigma_{\phi_{pq}=180}] /
[\sigma_{\phi_{pq}=0} + \sigma_{\phi_{pq}=180}]$, which exhibits
sensitivity to the Coulomb quadrupole amplitude, was also
determined. Measurements of the parallel cross section $\sigma_{0}$
were also performed in the range of W=1170~MeV to 1232~MeV. A first
level of acceptance cuts was applied in the data analysis in order
to limit the phase space to the central region of the spectrometers
and to ensure that potential edge effects will be avoided. For the
pair of $\phi_{pq}^{*}= 0^\circ$ and $180^\circ$ measurements the
cross sections, responses, and asymmetries were obtained with the
phase space matched in (W,$Q^2$,$\theta_{pq}^{*}$). Point cross
sections were extracted from the finite acceptances by utilizing the
cross section calculations from various theoretical
models~\cite{dmt,kama,mai00,multi,said} in the Monte Carlo
simulation. Radiative corrections were applied to the data using a
Monte Carlo simulation~\cite{mceep}. The cross section systematic
uncertainties are of the order of $\pm 3\%$, dominating over the
better than $\pm 1\%$ statistical uncertainties. In the asymmetry
measurements the systematic uncertainties were further suppressed
through the cross section ratio, while an advantage is presented by
the fact that the electron spectrometer position and momentum
settings do not change during the asymmetry measurements. A detailed
description of the data analysis is presented
in~\cite{anez,blomberg}.

In Fig.~\ref{fig:resp} the experimental results for $\sigma_{LT}$
and $\sigma_{o}$+$\epsilon{\cdot}\sigma_{TT}$ are presented, and in
Fig.~\ref{fig:asym} the asymmetry measurements are exhibited. In
Fig.~\ref{fig:asym} the measurement of the parallel cross section
$\sigma_{o}$ at $Q^2 = 0.13~(GeV/c)^2$ as a function of W is also
presented. The experimental results are compared with the SAID
multipole analysis \cite{said}, the phenomenological model MAID 2007
\cite{mai00,kama} and the dynamical model calculations of Sato-Lee
\cite{sato_lee} and of Dubna - Mainz - Taipei (DMT) \cite{dmt}. The
Sato-Lee \cite{sato_lee} and DMT \cite{dmt} are dynamical reaction
models which include pionic cloud effects. Both calculate the
resonant channels from dynamical equations. DMT uses the background
amplitudes of MAID with some small modifications. Sato-Lee calculate
all amplitudes consistently within the same framework with only
three free parameters. Both find that a large fraction of the
quadrupole multipole strength arises due to the pionic cloud with
the effect reaching a maximum value in this momentum transfer
region. Sato-Lee exhibits a relatively good agreement with the
$\sigma_{LT}$ measurements as one moves to lower $Q^2$ while DMT
systematically overestimates this response indicating an
overestimation of the Coulomb quadrupole amplitude. Both
calculations provide a reasonable agreement to the $\sigma_{o}$
measurements as a function of W as shown in Fig.~\ref{fig:asym}. On
the other hand the MAID model \cite{kama,mai00} which offers a
flexible phenomenology, as well as the SAID multipole analysis, fail
to reproduce the W-dependence of the $\sigma_{o}$ measurements. This
observation is in agreement with previous
measurements~\cite{longpaper,sparepja} that suggest that both
calculations need to be refined, especially at the lower wing of the
resonance. Both calculations perform reasonably well at the higher
$Q^2$ measurements but their predictions deviate more as one moves
lower in $Q^2$, as indicated by Fig.~\ref{fig:resp} and
Fig.~\ref{fig:asym}.

\begin{figure}[!]
\center
\includegraphics[width = \columnwidth]{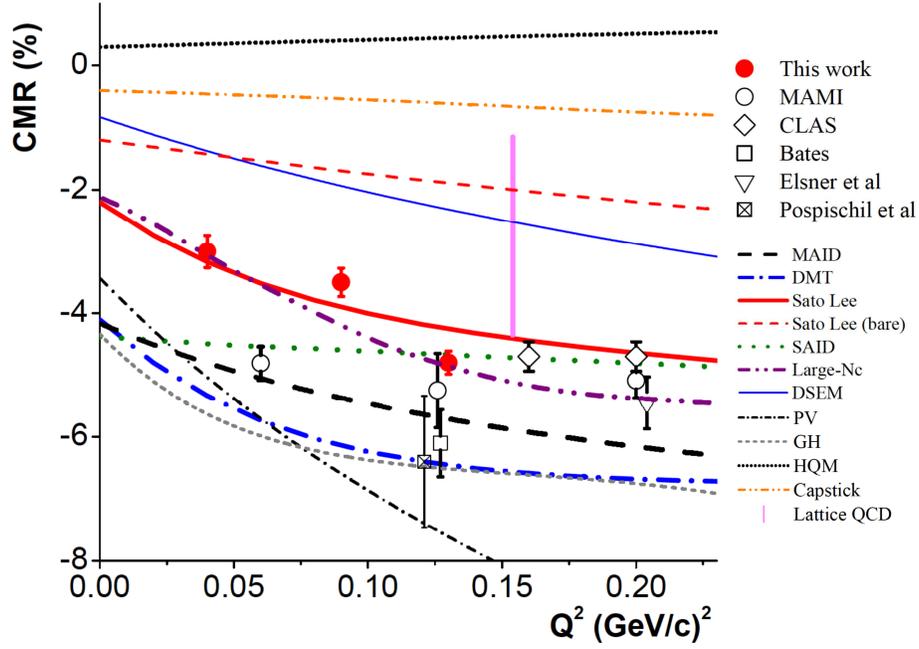}
\caption{The CMR measurements as a function of $Q^2$. The results
from this work (solid circles) and from
\cite{pos01,spaprl,stave,elsner,spaplb,aznauryan} (open symbols) are
presented. All data points are shown with their total experimental
 uncertainties (statistical and systematic) added in quadrature. The theoretical
predictions of MAID \cite{kama,mai00}, DMT \cite{dmt}, SAID
\cite{said}, Sato-Lee \cite{sato_lee}, Capstick \cite{capstick}, HQM
\cite{hqm}, the Lattice-QCD calculation \cite{dina}, the large-Nc
calculation \cite{largenc}, the DSEM~\cite{dsem}, the ChEFT  of
Pascalutsa-Vanderhaegen (PV)~\cite{pasc} and the Gail-Hemmert (GH)
\cite{hemmert} are also shown.} \label{fig:cmr}
\end{figure}

Fits of the resonant amplitudes have been performed while taking
into account the contributions of background amplitudes from the
MAID, DMT, SAID, and Sato Lee models. The fitting procedure is
described in detail in~\cite{blomberg} and it is the same that has
been applied before in various
experiments~\cite{spaplb,longpaper,sparepja}. The resonant
amplitudes are fitted while utilizing the background amplitudes from
each theoretical model calculation separately. The models differ in
their description of the background terms thus leading to a
deviation of the fitted results which indicates the level of the
model uncertainty. The deviation of the fitted central values is
adopted as a model uncertainty of the extracted amplitudes. For the
CMR ratio, at $Q^2=0.13~(GeV/c)^2$ we find a value of $(-4.80 \pm
0.19_{\text{stat+sys}} \pm 0.80_{\text{model}})\%$ which is in
excellent agreement with the recent MAMI
measurement~\cite{sparepja}. For $Q^2 = 0.09~(GeV/c)^2$ and $Q^2 =
0.04~(GeV/c)^2$ we find that the CMR is $(-3.50 \pm
0.20_{\text{stat+sys}} \pm 0.80_{\text{model}})\%$ and $(-3.00 \pm
0.27_{\text{stat+sys}} \pm 0.80_{\text{model}})\%$ respectively. The
EMR results, $(-2.50 \pm 0.50_{\text{stat+sys}} \pm
0.50_{\text{model}})\%$ at $Q^2 = 0.13~(GeV/c)^2$ and $(-1.90 \pm
0.50_{\text{stat+sys}} \pm 0.50_{\text{model}})\%$ at $Q^2 =
0.09~(GeV/c)^2$, confirm earlier measurements~\cite{longpaper} that
indicate that the ratio stays within 2$\%$-2.5$\%$ in this region.
The derived CMR values are presented in Fig.~\ref{fig:cmr}. One can
observe a disagreement between the MAMI result at $Q^2 =
0.06~(GeV/c)^2$~\cite{stave} and the new data. The source of this
disagreement has been identified in the extraction
procedure~\cite{stave} of the resonant amplitudes from the measured
MAMI cross sections. A revised extraction procedure corrects the CMR
value at $Q^2 = 0.06~(GeV/c)^2$~\cite{stave}, moving it towards the
new data by approximately 1$\%$ thus reconciling this discrepancy;
details of this revised work will be presented in an upcoming
publication.

As exhibited in Fig.~\ref{fig:cmr} the Sato Lee prediction has a
remarkable success in describing the $Q^2$ evolution of the Coulomb
quadrupole amplitude. The DMT, MAID, and SAID calculations are less
effective and tend to overestimate the magnitude of the ratio. The
data provide a strong support to the interpretation within the Sato
Lee model that the $\Delta$ resonance consists of a bare quark-gluon
core and a pion cloud, and the large pion cloud contribution to CMR
can be seen by comparing the Sato Lee solid and dashed curves in
Fig.3. We further observe that the dashed curve of the Sato Lee
``bare" component is qualitatively similar to the prediction of a
model based on the Dyson-Schwinger Equation of QCD~\cite{dsem}
(DSEM). Since DSEM does not include the pion degree of freedom, the
agreement between the data and the Sato Lee prediction suggests a
possible link between the bare quark-core of a dynamical model and
the genuine QCD dynamics.

%\section{Outlook}

The new data have accessed a kinematic region where, for the first
time, a more drastic change of the CMR magnitude with $Q^2$ is
observed compared to the trend of the world data in the region
higher than $Q^2 = 0.1~(GeV/c)^2$. The results suggest that the
values of the CMR and EMR ratios converge as $Q^2\rightarrow$ 0.
This is well described by the Sato Lee model, and considering that
the bare CMR and EMR values of the model are equal at $Q^2=0$ due to
the use of the long-wave limit, the convergence of the CMR and EMR
ratios at $Q^2=0$ suggests that the meson cloud contribution to both
quadrupole amplitudes is similar as we enter the low $Q^2$ regime.

In Fig.~\ref{fig:cmr} one can also identify the success of the
large-Nc calculation~\cite{largenc} in the prediction of the CMR
ratio. Nevertheless the calculation underestimates the values the
magnetic dipole and of the quadrupole amplitudes by $\approx$~20\%
but this effect cancels out in the ratio. Constituent quark model
(CQM) predicitions are known to considerably deviate from the
experimental results. Two representative CQM calculations are shown
in Fig.~\ref{fig:cmr}, that of Capstick \cite{capstick} and of the
hypercentral quark model (HQM) \cite{hqm}, which fail to describe
the data. It demonstrates that the color hyperfine interaction is
inadequate to explain the effect at large distances. Chiral
effective field theoretical calculations \cite{pasc,hemmert} also
account for the magnitude of the effects giving further credence to
the dominance of the meson cloud at the low momentum transfer
region. Chiral perturbation theory offers the natural framework to
investigate the role of pionic contributions to the nucleon
structure where nucleon observables receive contribution from pion
loops, the ``pion cloud", but it has to be noted that such
contributions are in general not scale-independent~\cite{meiss} and
thus can not provide a model independent definition.

Lattice QCD results~\cite{dina} allow a comparison to experiment
with the chirally extrapolated \cite{pasc} values of CMR found to be
nonzero and negative in the low $Q^2$ region. Lattice QCD
calculations~\cite{dinaprivate} that utilize improved methods are
currently ongoing and will provide results at lower $Q^2$, with
reduced uncertainties, and with lighter quark masses of 180~MeV.
These calculations so far indicate~\cite{dinaprivate} that the
discrepancy between Lattice QCD and the data gets smaller as the
pion mass approaches the physical value. Lattice QCD calculations
with pion mass close to the physical are now within reach in the
near future. The new results provided by this experiment offer
important, high precision, benchmark quantities. Namely, at physical
value of the pion mass and after taking the continuum limit Lattice
QCD should reproduce the data, otherwise it cannot claim to predict
other quantities.

In conclusion, we have reported on new p$(e,e^\prime p)\pi^\circ$
measurements at the $\Delta^{+}(1232)$ resonance at the low momentum
transfer region where the mesonic cloud dynamics are predicted to be
dominant and appreciably changing with $Q^2$. The Coulomb quadrupole
amplitude measurements have been extended to a new lowest momentum
transfer, and a rapid fall-off of the magnitude of the CMR ratio
below $Q^2 = 0.1~(GeV/c)^2$ has been observed for the first time at
low $Q^2$. The reported measurements reveal, for the first time,
that the values of the two quadrupole amplitudes converge as
$Q^2\rightarrow0$. The measured resonant amplitudes are in
disagreement with the values predicted by quark models on account of
the noncentral color-hyperfine interaction. On the other hand, the
dominant role of the mesonic degrees of freedom has been
demonstrated at the large distance scale. The new data are described
with a remarkable success from a dynamical model that suggests that
more than half of the magnitude of the Coulomb quadrupole amplitude
is attributed to the mesonic cloud at low $Q^2$. The same conclusion
is being further supported by a Dyson-Schwinger calculation Equation
Model where the pion degrees of freedom are not included, and the
fact that it underestimates the data is a clear and important
indication of the dressed-quark component. The results are in
qualitative agreement with chiral perturbation theory calculations,
and they also provide important benchmark quantities for the Lattice
QCD calculations. Strong experimental constraints have been provided
to the theoretical calculations, thus offering the necessary input
that will allow their refinement and will resolve the theoretical
discrepancies.

%===============================================
%Acknowledgements
%===============================================
%\section*{Acknowledgements}

We would like to thank the JLab Hall A technical staff and
Accelerator Division for their outstanding support, as well as T.-S.
H. Lee, C. Alexandrou, C. Roberts, A. Bernstein, V. Pascalutsa and
M. Vanderhaeghen for the useful discussions and correspondence. This
work is supported by the National Science Foundation award
PHY-1305536 and the UK Science and Technology Facilities Council
(STFC 57071/1, STFC 50727/1).
%===============================================
%Refs
%===============================================

\section*{References}

\bibliographystyle{elsarticle-num}

\end{document}